\pdfoutput=1
\documentclass[journal,twoside,web]{ieeecolor}
\usepackage{tmi}
\usepackage{cite}
\usepackage{amsmath,amssymb,amsfonts}
\usepackage{graphicx}
\usepackage{textcomp}
\usepackage{url}
\usepackage{algorithm,algpseudocode}

\makeatletter
\newenvironment{algorithmbox}[1]{%
  \par\addvspace{0.5\baselineskip}%
  \noindent\rule{\linewidth}{0.8pt}\par\vspace{2pt}%
  \refstepcounter{algorithm}%
  \noindent\textbf{Algorithm~\thealgorithm} #1\par\vspace{2pt}%
  \noindent\rule{\linewidth}{0.4pt}\par\vspace{2pt}%
}{%
  \par\vspace{2pt}\noindent\rule{\linewidth}{0.8pt}\par\addvspace{0.5\baselineskip}%
}
\makeatother

\def\BibTeX{{\rm B\kern-.05em{\sc i\kern-.025em b}\kern-.08em
    T\kern-.1667em\lower.7ex\hbox{E}\kern-.125emX}}
\begin{document}
\bstctlcite{IEEEtran:IEEEtranBSTCTL}

\title{Spatiotemporal Adaptive Eigenspectrum Remapping in Transcranial Ultrasound Localization Microscopy}
\author{Stephen A. Lee, \IEEEmembership{Member, IEEE}, Isidora Jelenkovic, Maia V. Kokoeva, and Jean Provost, \IEEEmembership{Member, IEEE}
\thanks{This work has been submitted to the IEEE for possible publication. Copyright may be transferred without notice, after which this version may no longer be accessible.}
\thanks{
This work was supported in part by the Vanier-Banting Postdoctoral Fellowship (NSERC), the CIHR Postdoctoral Fellowship, the TransMedTech Living Lab Postdoctoral Scholarship, the IVADO, Canada Foundation for Innovation (38095), the CIHR (452530), New Frontiers in Research Fund (NFRFE-2018-01312), the NSERC (RGPIN-2019-04982), Fonds de Recherche de Quebec – Nature et technologies, the Quebec Bio-Imaging Network, CONAHCYT, the Insightec, the Healthy Brains Healthy Lives, and the Canada First Research Excellence Fund, and the National Institutes of Aging (1R01AG079894), and the CIHR grant (PJT-175333).}
\thanks{S. A. Lee is with the Department of Engineering Physics, Polytechnique Montreal, Montreal, QC H3T 0A3 CA (e-mail:stephen.lee@polymtl.ca). }
\thanks{I. Jelenkovic and M. V. Kokoeva are with the Department of Neuroscience, McGill University, Montreal, QC, H3T 0A3, CA}
\thanks{J. Provost is with the Department of Engineering Physics, Polytechnique Montreal, Montreal, QC H3T 0A3 CA and the Montreal Heart Institute, Montreal, H1T 1C8, QC, CA (e-mail:jean.provost@polymtl.ca).}}

\maketitle

\begin{abstract}
Clutter suppression is a critical preprocessing step in ultrasound blood flow imaging, yet conventional singular value decomposition (SVD) filtering depends on data-specific eigenvalue cutoffs that can remove slow flow along with tissue clutter. Here, we introduce adaptive eigenspectrum remapping (AER), a general framework in which clutter is suppressed by continuously remapping, rather than truncating, the eigenvalue spectrum of the spatiotemporal covariance, and we investigate zero-phase component analysis (ZCA) matrix whitening as an instance of AER for transcranial ultrasound localization microscopy (ULM). Rather than discarding eigenspaces, ZCA-based AER whitens the eigenvalue spectrum, reducing the dominance of coherent clutter, including skull reverberation, while retaining weak flow signals. We evaluate AER on both temporal and spatial covariance matrices in mouse brain datasets acquired with long temporal ensembles. Compared with SVD filtering, ZCA improves microbubble contrast and vascular image quality, increases trajectory density across low- and high-concentration regimes, and reveals more low-velocity trajectories in ULM velocity maps. Spatial AER further maintains reconstruction quality across extended ensembles and enables fuller single capillary reporter (SCaRe) reconstructions, including connected artery-to-capillary-to-vein tracks. These results support AER, here instantiated as ZCA whitening, as a data-adaptive alternative to manually thresholded SVD filtering for high-resolution microvascular imaging.
\end{abstract}

\begin{IEEEkeywords}
Ultrasound Localization Microscopy (ULM), Clutter Filtering, Adaptive Eigenspectrum Remapping (AER), Spatiotemporal Whitening, Zero-Phase Component Analysis (ZCA), Transcranial Ultrasound, Functional Ultrasound.
\end{IEEEkeywords}

\section{Introduction}
\label{sec:intro}
\IEEEPARstart{U}{ltrasound} localization microscopy (ULM) is an imaging methodology adapted from optical super-resolution techniques such as PALM and STORM \cite{almada2015palm}, in which microbubbles, used in Contrast-Enhanced Ultrasound (CEUS) at high concentrations, are digitally isolated and individually localized, enabling assessment of micron-scale angioarchitecture in whole organs \cite{couture2018ultrasound,wu20253d,leconte2024tracking,bourquin2024quantitative}. However, it is more clinically-relevant to go beyond static structural imaging and measure functional dynamics in the brain. Such applications are currently being developed: dynamic ULM (dULM) tracks blood flow over time to measure pulsatility \cite{wu20253d,bourquin2024quantitative} and functional ULM (fULM) interprets changes in cerebral blood flow as a correlate of neural activity \cite{renaudin2022functional,shin2026microbubble}. All of these techniques, however, depend critically the accurate delineation between clutter, stationary, and slow-moving tissue; this "clutter filtering problem" becomes exceedingly harder at the microvascular level, especially at the capillary scale \cite{lee2026assessing}.

Early clutter filtering methods assume blood and tissue occupied distinct, non-overlapping spectral profiles: tissue signal concentrated near the DC component while blood signal shifted higher in the Doppler spectrum, enabling the use of finite impulse response (FIR) and infinite impulse response (IIR) filters \cite{bjaerum2002clutter}. The advent of ultrafast ultrasound improved microvascular imaging performance by allowing denser temporal sampling \cite{tanter2014ultrafast}. Subsequently, in viewing ultrasound data as a mixture of spatiotemporal correlations (stationary tissue and clutter exhibit high spatiotemporal coherence, while blood signal does not \cite{bjaerum2002clutter}), a matrix decomposition technique called the Singular Value Decomposition (SVD) has been applied to separate flow based on eigensubspace characteristics \cite{baranger2018adaptive}. This approach, a generalization of the principal component analysis (PCA), has since been widely adopted for clutter filtering in CEUS and ULM \cite{demene2015spatiotemporal}.

Despite its robustness and subsequent extensions, such as higher-order SVD \cite{wildeboer2020blind} and multi-stage SVD \cite{zhang2027multistage}, SVD-based filtering still depends on accurately estimating coherent energy in eigenspace and discarding singular values corresponding to highly coherent clutter structures. This approach does not fully resolve the separation of capillary-level slow flows ($<$1 mm/s \cite{lee2026assessing}) from tissue signal, and requires either an automatic or manual eigenvalue cutoff \cite{baranger2018adaptive}. Nonetheless, the appropriate choice of cutoff has been widely debated \cite{riemer2023use} and its implementation remains inconsistent across the field, of which the selected value can substantially influence which vascular structures are recovered in ULM. Crucially, filter quality is difficult to judge from image quality alone, since a cutoff can be tuned to look visually convincing while still discarding functionally relevant flow overlapping with clutter. We therefore use our ability to resolve capillaries that depend on slow flow estimation as the functional benchmark for evaluating clutter filtering throughout this work.

This fundamental problem resembles a classical challenge in optimal filtering theory: detecting a known signal buried in colored noise, for which the optimal detector under stationary Gaussian noise is given by Wiener matching \cite{cuoco2004whitening}. The operation of converting to a flat power spectral density is called 'whitening': by whitening the spectrum so that all frequencies contribute equally, the filter attenuates dominant, predictable noise frequencies and amplifies the underlying "innovations" of the target signal, which in spatiotemporal ultrasound data corresponds to flow. In this work, we translate this principle to the spatiotemporal domain: if tissue clutter dominates the leading eigenvalues of the spatiotemporal autocovariance matrix, then whitening the eigenvalue spectrum suppresses clutter and amplifies the relative contribution of flow, without requiring an explicit threshold. 

We therefore introduce adaptive eigenspectrum remapping (AER), an umbrella term we use for the family of clutter filters that retain every eigenspace of the spatiotemporal covariance and instead apply a weight to each one that is a function of its own eigenvalue, so that the filter adapts to the measured eigenspectrum of each dataset. Conventional SVD thresholding can be viewed as the special case of binary (0/1) weights; AER instead favors continuous weightings. In this work, we use zero-phase component analysis (ZCA) matrix whitening \cite{kessy2018optimal} as our AER weighting to enhance microbubble reconstruction. We adopt the broader term deliberately: whitening is one principled choice of weights, but other weightings of the eigenvalues may perform better for particular acquisitions, anatomies, or flow regimes. Rather than discarding singular values, which irreversibly removes spatiotemporal information, ZCA-based AER whitens the coherence energy across all subspaces, decorrelating them and normalizing their variance while preserving the underlying statistical structure. Because tissue clutter possesses highly imbalanced spatiotemporal coherence energy relative to flow, this transformation reduces the dominance of coherent clutter and amplifies the relative contribution of flow signal. Crucially, the absence of a hard eigenvalue cutoff enables a smooth transition across subspaces, facilitating recovery of slow microbubble flows that would otherwise be discarded by conventional SVD thresholding. We then characterize how AER with ZCA weighting can be used to improve microbubble flow separation in the mouse brain for ULM tracking and capillary vessel reconstruction.

\section{Theory}
\label{Theory}

Here we detail the theory of AER as applied to spatiotemporal ultrasound data for enhancing microbubble reconstruction. We develop the framework through the lens of dimensionality reduction, beginning with the signal model and Casorati matrix formulation, proceeding through a spectral and covariance interpretation of the data, defining the general AER transform, then ending with the practical ZCA whitening weighting of the covariance matrix used throughout this work.

\subsection{Ultrasound Signal Model}

We define our signal of interest for 2D ultrasound (though the framework can extend directly into 3D space):

\begin{equation}
    u(x,z,t) = s(x,z,t) + v(x,z,t) + \delta(x,z,t)
\end{equation}

where $u$ is the beamformed in-phase quadrature (IQ) ultrasound signal, $s$ is the stationary or slowly-varying clutter signal (e.g., tissue or skull reverberation), $v$ is the dynamic signal of interest (e.g., blood flow or microbubble contrast agents), and $\delta$ is system noise. The classical clutter filtering problem is thus to recover $v$ by suppressing $s$. However, posing this problem as such may be an oversimplification of the reality: in practice, tissue and flow signals can exhibit overlapping spatiotemporal dynamics, making removal of unwanted clutter non-trivial.

Nonetheless, a robust method to tackle this problem is to exploit the spatiotemporal correlations of the data. This can be performed by reshaping the ultrasound data into a Casorati matrix formulation, where the spatial (pixel) dimensions $p = (x,z)$ are vectorized into a single index of length $N_p = N_x \times N_z$:

\begin{equation}
    \mathbf{U} \in \mathbb{C}^{N_p \times N_t}
\end{equation}

\begin{equation}
\mathbf{U} = 
\begin{bmatrix}
u(p_1, t_1) & u(p_1, t_2) & \cdots & u(p_1, t_{N_t}) \\
u(p_2, t_1) & u(p_2, t_2) & \cdots & u(p_2, t_{N_t}) \\
\vdots       & \vdots       & \ddots & \vdots            \\
u(p_{N_p}, t_1) & u(p_{N_p}, t_2) & \cdots & u(p_{N_p}, t_{N_t})
\end{bmatrix}
\end{equation}

where each column $\mathbf{U}(:, t_k)$ represents a full spatial frame at observed time $t_k$, and each row $\mathbf{U}(p_i, :)$ is the temporal signal at pixel $p_i$. Clutter filtering can be performed directly on $\mathbf{U}$; however, a more principled approach is to operate on the second-order statistics, i.e., the covariance matrix. Often this will be the temporal covariance matrix since it is more computationally- and memory-efficient. For the mean-subtracted (centered) Casorati matrix $\mathbf{\tilde{U}}$, this is:

\begin{equation}
    \mathbf{C} = \frac{1}{N_t - 1} \mathbf{\tilde{U}}^H \mathbf{\tilde{U}} 
    \in \mathbb{C}^{N_t \times N_t}
\end{equation}

$\mathbf{C}$ is Hermitian positive semi-definite with rank at most $\min(N_p, N_t)$, and encodes the temporal correlations between all pairs of frames, averaged across all spatial locations.

\subsection{Spectral Analysis and the Wiener-Khinchin Theorem}

An initial approach to clutter filtering is to analyze the spectral content of the signal and suppress frequencies associated with stationary structures. The Wiener-Khinchin theorem \cite{papoulis2002probability} provides the theoretical groundwork, stating that the power spectral density (PSD) $S(f)$ is the Fourier transform of the autocorrelation function $R(\tau)$:

\begin{equation}
    S(f) = \mathcal{F}\{R(\tau)\} = \int_{-\infty}^{\infty} R(\tau) \, 
    e^{-i 2\pi f \tau} \, d\tau
\end{equation}

where the autocorrelation $R(\tau) = \mathbb{E}[u(t)u^*(t+\tau)]$ describes temporal self-similarity at lag $\tau$. In the discrete, finite-dimensional setting, a spatially averaged temporal covariance matrix can be related to a discretized autocorrelation matrix. Under additional periodic or circulant boundary assumptions, the discrete Fourier basis diagonalizes this covariance and its eigenvalues correspond to sampled power spectral densities. Real ultrasound data, however, is a finite recording whose statistics drift over time, rarely having a clean repeating structure. That mismatch is exactly why a filtering basis that adapts to the data itself, rather than one built from a fixed set of frequencies, is useful here \cite{taylor2019ensemble}.

\subsection{Principal Component Analysis and the SVD}

Principal Component Analysis (PCA) provides a data-adaptive alternative to Fourier decomposition. It decomposes data into an orthogonal basis that sequentially maximizes explained variance, via the eigendecomposition of the temporal covariance matrix:

\begin{equation}
    \mathbf{C} = \mathbf{V} \mathbf{\Lambda} \mathbf{V}^H
\end{equation}

where $\mathbf{V} = [v_1, v_2, \dots, v_{N_t}]$ are the eigenvectors (temporal principal components) and $\mathbf{\Lambda} = \mathrm{diag}(\lambda_1 \geq \lambda_2 \geq \cdots \geq \lambda_{N_t} \geq 0)$ are the corresponding eigenvalues. Each $\lambda_k$ represents the variance explained by the $k$-th mode, with total variance $\mathrm{tr}(\mathbf{C}) = \sum_k \lambda_k$.

The theoretical foundation for PCA is the \textit{Karhunen-Lo\`{e}ve Transform} (KLT) \cite{fukunaga1990introduction}, which states that for a continuous stochastic process, the eigenfunctions of the covariance kernel form the optimal orthogonal basis by minimizing mean-squared reconstruction error for any fixed number of retained components. The KLT can be understood as a data-adaptive counterpart to the DFT: whereas the DFT projects onto a fixed sinusoidal basis, the KLT derives its basis directly from the data. They coincide under the appropriate stationary and periodic, or circulant, assumptions \cite{dixon1993detection}. Thus, PCA on the spatially averaged temporal covariance provides a finite-dimensional analogue to Fourier spectral analysis while remaining applicable when ultrasound statistics are nonstationary or spectrally overlapping.

PCA can equivalently be computed from the SVD of the spatially averaged temporal covariance matrix $\mathbf{C}$. Because $\mathbf{C}$ is Hermitian positive semi-definite, this SVD is identical to the eigendecomposition above: both singular-vector matrices are $\mathbf{V}$, and the singular values equal the eigenvalues ($\mathbf{\Sigma} = \mathbf{\Lambda}$). To recover the spatial structure of each mode (i.e., image reconstruction), the temporal eigenvectors are projected back onto the Casorati matrix:

\begin{equation}
    \mathbf{\Phi} = \mathbf{\tilde{U}}\mathbf{V} \in \mathbb{R}^{N_p \times N_t}
\end{equation}

where each column $\mathbf{\Phi}(:,k) = \mathbf{\tilde{U}}v_k$ is the spatial map of the $k$-th temporal mode. The filtered data are then reconstructed by retaining only the subset $\mathcal{S}_v$ of modes associated with flow:

\begin{equation}
    \mathbf{\tilde{U}}_{\text{filtered}} = \mathbf{\Phi}_v \mathbf{V}_v^H = 
    \mathbf{\tilde{U}}\mathbf{V}_v\mathbf{V}_v^H
\end{equation}

The SVD of $\mathbf{C}$ can therefore be interpreted as a data-adaptive spectral decomposition: the eigenvectors form an empirically derived basis that diagonalizes $\mathbf{C}$, with eigenvalues $\lambda_k$ encoding the temporal variance of each mode. Standard SVD-based clutter filtering exploits the commonly observed separation between high-variance clutter modes and lower-variance flow modes by thresholding, retaining selected flow-associated eigenvectors and discarding the rest via the orthogonal projection $\mathbf{\tilde{U}}_{\text{filtered}} = \mathbf{\tilde{U}}\mathbf{V}_v\mathbf{V}_v^H$.

However, this threshold is inherently problem-dependent: the boundary between clutter and flow in the eigenvalue spectrum shifts with imaging conditions, acquisition parameters, depth, contrast agent concentration, and tissue motion. As a result, a fixed threshold generalizes poorly across datasets, and the absolute energetic representation of flow signals can vary substantially between experiments and systems, leading to the inconsistent implementation of SVD eigenvalue cutoffs. Furthermore, hard thresholding discards clutter components entirely, which risks removing signal energy in cases where clutter and flow subspaces overlap as well as lower image quality.

This motivates a complementary approach: rather than selecting which components to remove, can we \textit{transform} the eigenvalue spectrum itself? Specifically, if we whiten all eigenvalues to unit variance prior to filtering, the energetic dominance of clutter is suppressed and the relative contribution of flow modes is amplified, all without discarding any data or requiring an explicit threshold.

\subsection{Adaptive Eigenspectrum Remapping}

We formalize this idea as adaptive eigenspectrum remapping (AER). Given the eigendecomposition $\mathbf{C} = \mathbf{V}\mathbf{\Lambda}\mathbf{V}^H$, an AER transform applies a scalar weight $g(\lambda_k)$ to each eigenmode and rotates back into the original data space:
\begin{equation}
    \label{eq:aer}
    \mathbf{W}_g = \mathbf{V}\, g(\mathbf{\Lambda})\, \mathbf{V}^H, \qquad g(\mathbf{\Lambda}) = \mathrm{diag}\big(g(\lambda_1), \dots, g(\lambda_{N_t})\big)
\end{equation}
The transform is \textit{adaptive} because the weights are determined by the measured eigenspectrum of each dataset rather than by a fixed spectral or rank boundary. Hard-threshold SVD filtering is recovered as the special case $g(\lambda_k) \in \{0, 1\}$, which yields the orthogonal projection $\mathbf{V}_v\mathbf{V}_v^H$. AER instead favors continuous weightings that rescale, rather than discard, eigenmodes. Any weighting that decreases with $\lambda_k$ compresses the dynamic range of the eigenspectrum and therefore reduces the dominance of high-energy coherent clutter. In this work we use the whitening weighting $g(\lambda_k) = \lambda_k^{-1/2}$ (ZCA, below), which flattens the spectrum completely. We refer to the broader family deliberately, because other weightings, such as partial whitening $g(\lambda_k) = \lambda_k^{-\alpha}$ with $0 < \alpha < 1/2$, shrinkage-type weights, or weights informed by the temporal frequency content of each eigenvector, may provide a better balance between clutter suppression and noise amplification in some cases.

\subsection{Matrix Whitening}

Matrix whitening, the AER weighting used in this work, is a linear transformation that decorrelates the data and, in the unregularized case, normalizes each component to unit variance \cite{kessy2018optimal}. Whitening can take many forms. In the Fourier domain, for example, the data may be weighted by the inverse square root of the noise power spectral density, $d_\mathrm{w}(f) \propto d(f)/[S_n(f)]^{1/2}$ \cite{cuoco2004whitening}, so that comparisons across frequencies emphasize signal innovations. Here, the analogous goal is to reduce the dominance of highly coherent tissue clutter relative to weaker flow and noise. Given a full-rank temporal covariance matrix $\mathbf{C}$, an exact whitening transformation $\mathbf{W}$ satisfies:

\begin{equation}
    \mathbf{W} \mathbf{C} \mathbf{W}^H = \mathbf{I}
\end{equation}

This maps the data into a space where the covariance is approximately the identity, with temporal modes uncorrelated and carrying comparable variance. This can be seen in Fig.~\ref{fig:qualitative_covariance}. Geometrically speaking, whitening rescales the data distribution toward a sphere, hence the equivalent term \textit{sphering}. In practice, regularized whitening attenuates high-variance coherent modes and increases the relative visibility of weaker modes, although it may also amplify noise or weak clutter components.

The whitening matrix, however, is not unique; any valid $\mathbf{W}$ can be left-multiplied by an arbitrary unitary rotation $\mathbf{Q}$ ($\mathbf{Q}\mathbf{Q}^H = \mathbf{I}$) and remain a valid whitening matrix:

\begin{equation}
    \label{eq:identity}
\begin{aligned}
    \mathbf{W}^* &= \mathbf{Q}\mathbf{W} \\
    \mathbf{W}^*\mathbf{C}\mathbf{W}^{*H} &= \mathbf{Q}\mathbf{W}\mathbf{C}\mathbf{W}^H\mathbf{Q}^H \\
                                           &= \mathbf{Q}\mathbf{I}\mathbf{Q}^H \\
                                           &= \mathbf{I}
\end{aligned}
\end{equation}

Thus, the family of valid whitening matrices is parameterized by the choice of $\mathbf{Q}$. Kessy, Lewin, and Strimmer \cite{kessy2018optimal} showed that there exist at least five distinct whitening transformations, each optimal under different criteria relating the whitened data to the original. ZCA, ZCA-correlation, and Cholesky transformations all involve a rotation back into the original data space, thereby preserving spatiotemporal interpretability and flow visualization; we focus here only on ZCA. PCA-based whitening, in contrast, projects onto the eigenspace of $\mathbf{C}$ without rotating back, destroying the correspondence between whitened components and original spatial locations, a property that makes it less useful for medical imaging.

\subsubsection{ZCA Whitening}

Using the eigendecomposition $\mathbf{C} = \mathbf{V}\mathbf{\Lambda}\mathbf{V}^H$, the ZCA (Zero-phase Component Analysis) whitening matrix for a full-rank covariance is:

\begin{equation}
    \mathbf{W}_{\text{ZCA}} = \mathbf{V} \mathbf{\Lambda}^{-1/2} \mathbf{V}^H 
    = \mathbf{C}^{-1/2}
\end{equation}

This is the AER transform of Equation~\ref{eq:aer} with $g(\lambda_k) = \lambda_k^{-1/2}$: ZCA first rotates the data into the PCA basis via $\mathbf{V}^H$, rescales each component by $\lambda_k^{-1/2}$, then rotates back via $\mathbf{V}$. In the unregularized full-rank case, ZCA is the whitening transform closest to the original data in the least-squares sense \cite{kessy2018optimal}, preserving spatial structure while whitening the eigenvalue spectrum.

%
%
%
%
%

%
%
%

\subsubsection{Regularization}
Since rescaling by $\lambda_k^{-1/2}$ can be unstable for small eigenvalues, we use Tikhonov regularization to stabilize the transform:
\begin{equation}
    \quad \mathbf{C}_{\text{reg}} = \mathbf{C} + \lambda \mathbf{I}
\end{equation}

Here $\lambda$ controls the degree of regularization and whitening. The resulting regularized ZCA transform is:
\begin{equation}
    \label{eq:reg}
    \mathbf{W} = \mathbf{V}(\mathbf{\Lambda} + \lambda \mathbf{I})^{-1/2}\mathbf{V}^H
\end{equation}

\subsection{Why AER enhances microbubble reconstruction}

ZCA whitening operates on the eigenvalue spectrum of $\mathbf{C}$: it maps $\mathbf{\Lambda} \rightarrow \mathbf{I}$, collapsing all eigenvalues to unity regardless of their original magnitude. But why does this enhance microbubble reconstruction? Consider the structure of the raw eigenvalue spectrum: clutter signals are spatially coherent and temporally stationary, producing a small number of very large eigenvalues that dominate the spectrum. Flow signals, being less coherent and faster moving, are broadly distributed across many smaller eigenvalues that are effectively masked by the clutter energy. This whitening removes the energetic dominance of clutter, elevating the relative contribution of flow modes and suppressing its influence without having to discard any components or requiring a manual threshold. The same reasoning applies to any AER weighting that decreases with $\lambda_k$; whitening is the limiting case in which the eigenspectrum is flattened completely.

\begin{figure}[!t]
    \centering
    \includegraphics[width=\linewidth]{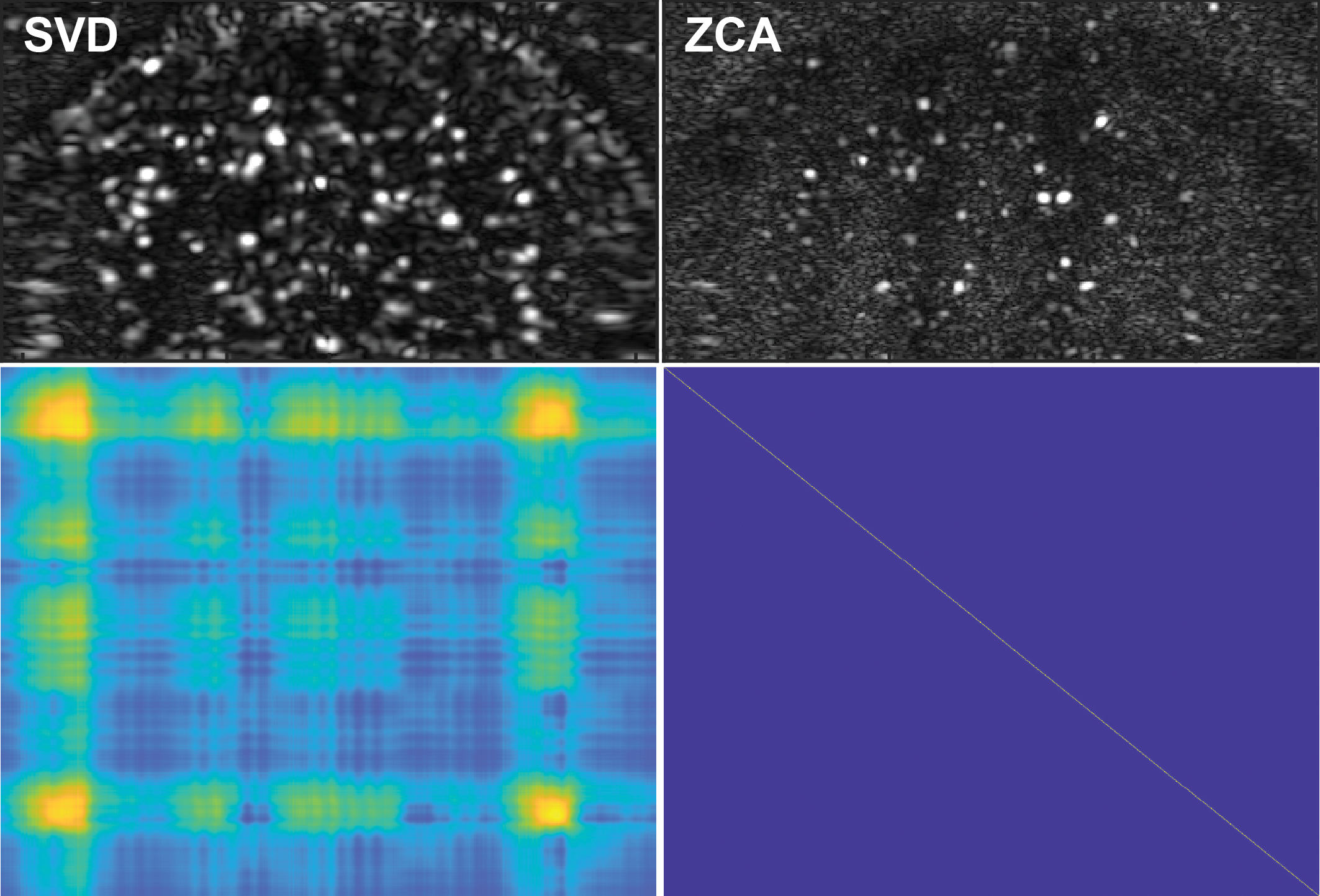}
    \caption{Qualitative comparison between singular value decomposition filtered data and AER with ZCA whitening. (top) A single frame of microbubble data flowing in the brain acquired transcranially through intact skin and skull, SVD filtered is on the left and ZCA whitening is on the right. (bottom) covariance of the temporal Casorati matrix composed from original RF data and RF data after whitening matrix projection.}
    \label{fig:qualitative_covariance}
\end{figure}

\section{Methods}
\label{Methods}

\begin{figure*}[!t]
    \centering
    \includegraphics[width=\linewidth]{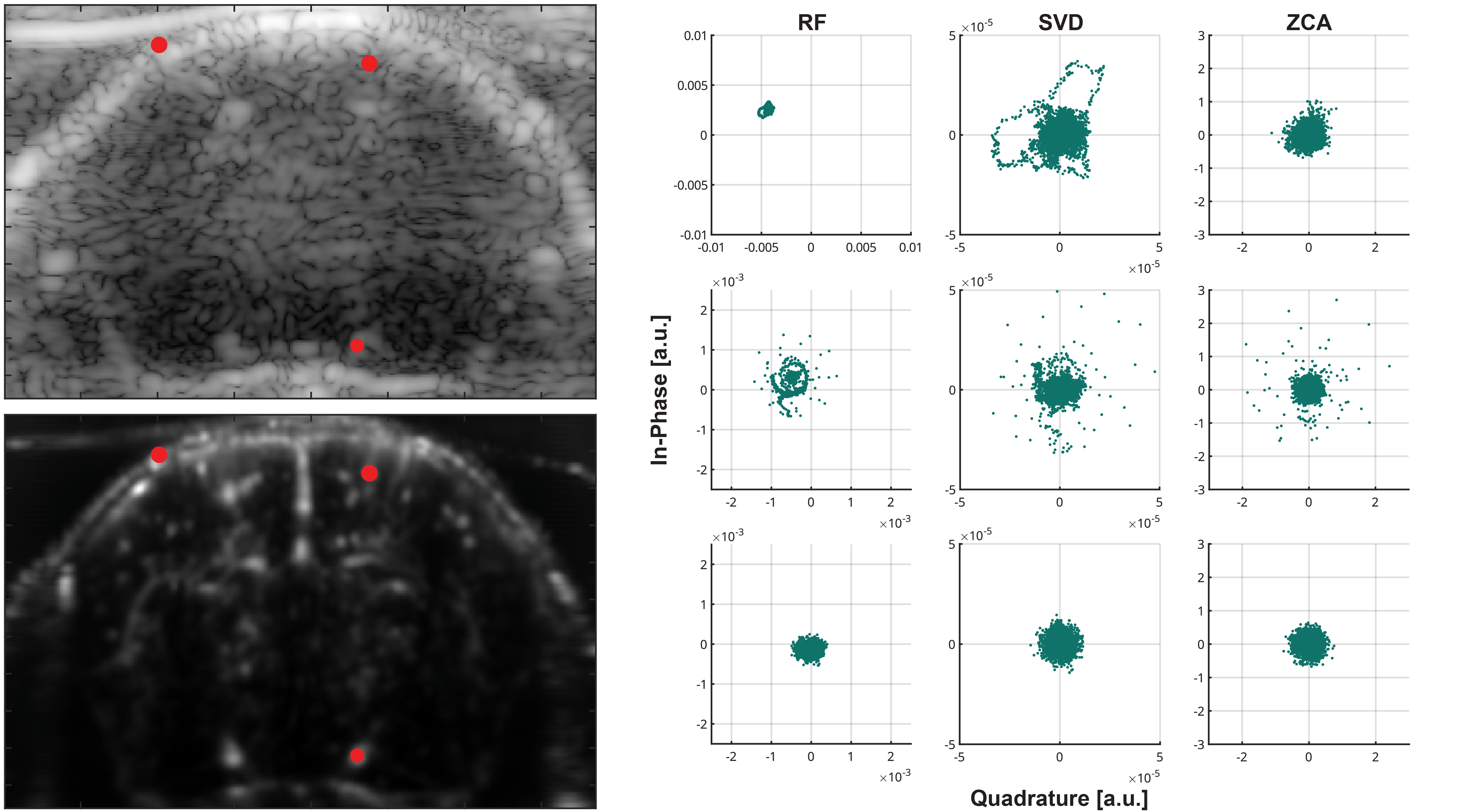}
    \caption{Exploring spatiotemporal features using IQ plots. Three points were chosen (skull, large cortical vessel, and large sub cortical vessel) and overlaid on unfiltered (top) and filtered (bottom) images of the brain. IQ plots for each point corresponds to each row and each column represents a different processing condition.}
    \label{fig:real_imag}
\end{figure*}

\subsection{Ultrasound Data}
Ultrasound ensembles were acquired in $n=10$ wild-type C57BL6 mice (male, 6–8 weeks old) under ketamine/xylazine sedation with continuous Sonovue microbubble infusion (15~$\mu$L/min). All animal procedures and husbandry were carried out in accordance with the recommendations of the Canadian Council on Animal Care (CCAC) and have been approved by the local McGill University Animal Care Committee. IQ data at 50\% bandwidth sampling were recorded on a Verasonics Vantage \& NXT scanner with gapless acquisition, using \textit{int16} and \textit{half} datatypes across datasets. Beamformed images were coherently compounded from 11 tilted plane waves ($-5^\circ$ to $5^\circ$) at a PRF of 1000~Hz, acquired at 15.6~MHz with a 128-element linear array (L22-14v, Vermon). Sessions lasted 30--40 minutes, yielding $\sim$2~TB per session. Images were beamformed via GPU-accelerated delay-and-sum (DAS).

\subsection{Spatiotemporal AER}
Ultrasound data were arranged into a Casorati matrix $[N_p, N_t]$, and the autocovariance matrix $C = \text{cov}(U)$ was computed on mean-subtracted data with Tikhonov regularization applied to its diagonal. AER was implemented with the ZCA weighting: the SVD of the regularized covariance yielded eigenvectors $U$ and eigenvalues $\Lambda$ used to construct the whitening matrix $W$. Remapped data were then recovered via:
\begin{equation}
    \tilde{U}_{\text{filtered}} = \tilde{U} \cdot W^H
\end{equation}

Algorithm~\ref{alg:zca} summarizes AER with ZCA weighting applied to the temporal covariance of a spatiotemporal (Casorati) ultrasound buffer. Other AER weightings are obtained by replacing the weight $s_k$ in line 5.

\begin{algorithmbox}{AER (ZCA Weighting) of Temporal Covariance}
\label{alg:zca}
\begin{algorithmic}[1]
\Require Casorati matrix $\tilde{U} \in \mathbb{R}^{N_p \times N_t}$, regularizer $\lambda$
\Ensure Whitened data $\tilde{U}_{\text{filtered}}$
\State $\tilde{U} \gets \tilde{U} - \text{mean}(\tilde{U}, \text{dim}=2)$ \Comment{mean-subtract}
\State $C \gets \tilde{U}^H \tilde{U} / N_p$ \Comment{temporal covariance}
\State $(U, \Lambda) \gets \text{SVD}(C)$ \Comment{eigendecomposition}
\For{$k = 1$ \textbf{to} $N_t$}
    \State $s_k \gets 1 / \sqrt{\lambda_k + \lambda}$ \Comment{AER weight $g(\lambda_k)$: ZCA}
\EndFor
\State $W \gets U \, \text{diag}(s) \, U^H$
\State $\tilde{U}_{\text{filtered}} \gets \tilde{U} W^H$
\end{algorithmic}
\end{algorithmbox}

\subsubsection{AER of the Temporal Covariance}

AER can be applied to any formulation of the covariance matrix. For the temporal case, we first arrange the data into the Casorati formulation, with space along one dimension and time along the other, then construct the temporal covariance matrix by transposition and multiplication. The whitening matrix is obtained by taking the SVD of this covariance matrix and rescaling its eigenvalues accordingly, achieving temporal whitening. Because data were acquired gaplessly, an arbitrary number of temporal samples can be incorporated; doing so, however, grows the size of the covariance matrix and increases computational complexity.

\begin{figure}[!t]
    \centering
    \includegraphics[width=\linewidth]{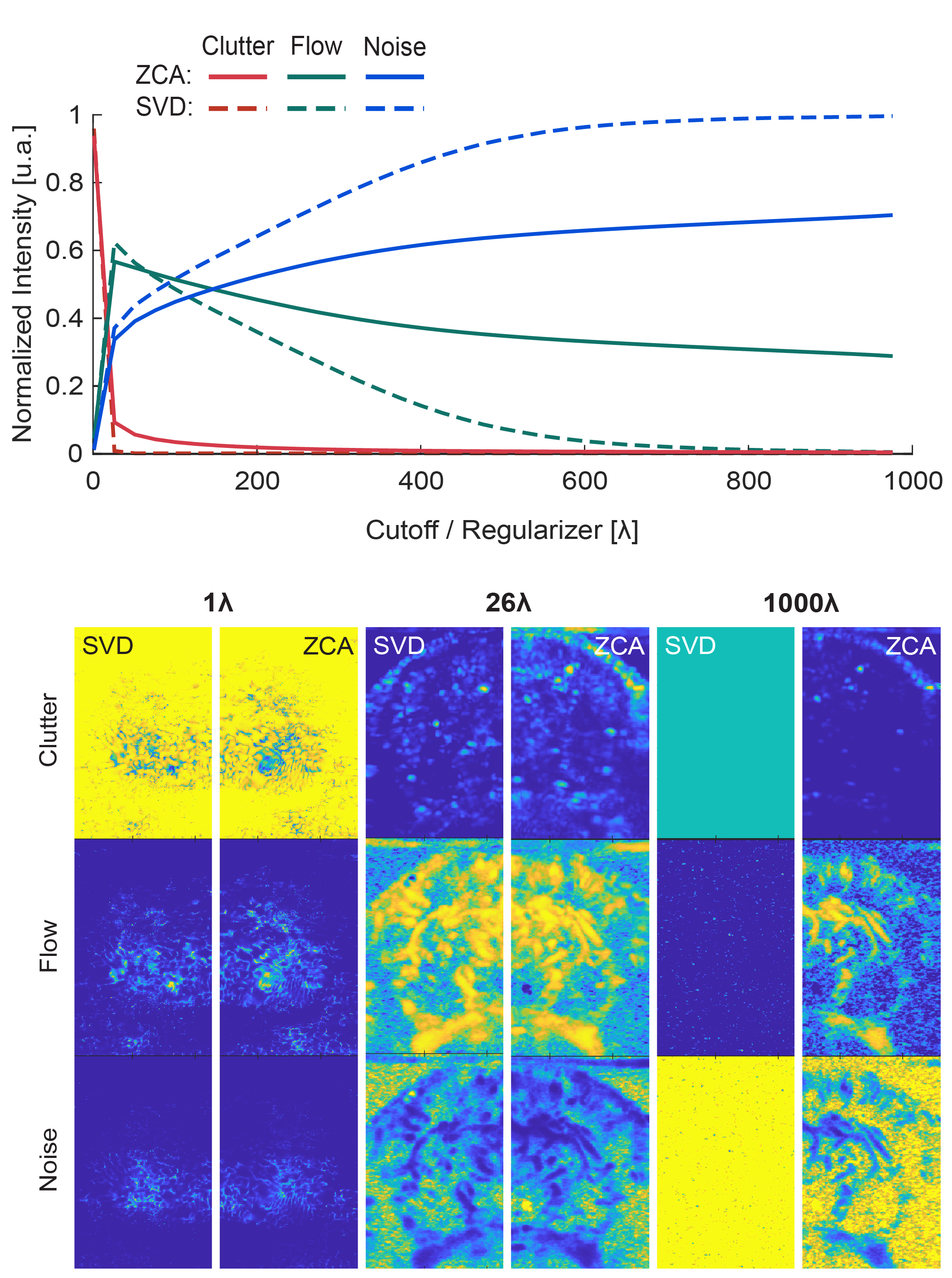}
    \caption{Spatiotemporal signal changes as a function of the regularization parameter $\lambda$. (top) Normalized power for clutter, flow, and noise as a function of $\lambda$. (bottom) Representative reconstructions for each Fourier band at $\lambda = 1$, $26$, and $1000$.}
    \label{fig:ens_compare}
\end{figure}

\subsubsection{AER of the Spatial Covariance}

Spatial AER follows a similar procedure to temporal AER, but operates on the spatial covariance matrix instead. Because the spatial dimensions do not grow with time, computational complexity remains more manageable than for temporal operations, given that the number of spatial samples stays fixed. Here, the spatial covariance was constructed iteratively by accumulating outer products of spatial vectors from each time frame, then normalizing by the number of frames, as shown in Algorithm~\ref{alg:spatial_cov}. This running accumulation lets the spatial covariance be estimated from arbitrarily long temporal ensembles without holding every frame in memory at once.

\begin{algorithmbox}{Iterative Accumulation of Spatial Covariance}
\label{alg:spatial_cov}
\begin{algorithmic}[1]
\Require Sequence of spatial buffers $X_1, \dots, X_K$, each $X_k \in \mathbb{R}^{N_p \times N_k}$
\Ensure Spatial covariance $C_p \in \mathbb{R}^{N_p \times N_p}$
\State $n \gets 0$, $s_x \gets \mathbf{0}_{N_p}$, $s_{xx} \gets \mathbf{0}_{N_p \times N_p}$
\For{$k = 1$ \textbf{to} $K$}
    \State $n \gets n + N_k$ \Comment{accumulate frame count}
    \State $s_x \gets s_x + \sum_j X_k(:,j)$ \Comment{accumulate spatial sum}
    \State $s_{xx} \gets s_{xx} + X_k X_k^H$ \Comment{accumulate outer-product sum}
\EndFor
\State $\bar{x} \gets s_x / n$
\State $C_p \gets (s_{xx} - n\, \bar{x}\bar{x}^H) / (n-1)$
\end{algorithmic}
\end{algorithmbox}

\subsection{Localization and Tracking}
\label{sec:tracking}

Two localization configurations were used, depending on the covariance formulation being evaluated. For the whole-brain temporal covariance comparisons, tracking was performed using the complete tracking-prior-to-localization workflow of Leconte et al. \cite{leconte2024tracking}, applied identically to SVD- and ZCA-filtered data. For the long-ensemble spatial covariance comparisons, this full workflow was not used, as its centerline-thinning step failed to form complete capillary tracks on these data. Instead, only the Hessian-based vesselness filter from \cite{leconte2024tracking} was retained to generate an approximate \textit{a priori} confidence map of bubble location from the lag-1 correlation image. Candidate bubble positions were localized as local maxima of the radial-symmetry response within this map, and candidates were linked across frames using the Hungarian algorithm \cite{munkres1957algorithms}. This reduced pipeline compensated for frame gaps and recovered additional stationary bubbles that were otherwise discarded by the full workflow.

\subsection{Parameter Optimization}
\label{sec:param_opt}

The regularization parameter $\lambda$ was swept from 1 to 1000 to characterize clutter, flow, and noise power tradeoffs (Section~\ref{sec:results}), and from 1 to 80 for capillary-count optimization on the subcortical ROI. For each value of $\lambda$, clutter, flow, and noise power were estimated by taking the temporal Fourier transform of the filtered signal at each pixel, spanning the full, unfolded 0--1000~Hz range set by the 1000~Hz PRF (IQ data have no Nyquist fold), and computing the mean power within a corresponding frequency band: clutter as 0--1~Hz, flow as 1--500~Hz, and noise as $>$500~Hz. The clutter cutoff (1~Hz) was set empirically as the lowest boundary that isolated stationary tissue without including slow flow (a 5~Hz cutoff was also tested but included visibly moving structures); the noise cutoff (500~Hz) was set at half the frame rate. The same band definitions were used for the capillary-count optimization sweep. Temporal ensemble size for the spatial-covariance analysis was optimized by measuring the effective rank of the eigenvalue spectrum of the spatial covariance matrix, defined as the participation ratio
\begin{equation}
    R = \frac{\left(\sum_k \lambda_k\right)^2}{\sum_k \lambda_k^2}.
\end{equation}
This measure quantifies the effective number of eigenmodes contributing comparable variance to the spectrum, i.e., how concentrated or spread the eigenvalue distribution is, rather than a literal count of independent flow components. An ensemble length of 15~s was selected from the plateau in $R$ as a function of ensemble size and used for all subsequent spatial AER comparisons.

\subsection{Single Capillary Reporters (SCaRe)}
\label{sec:scare}

Capillary-level flow recovery was quantified using the Single Capillary Reporter (SCaRe) framework \cite{lee2026assessing}, which uses a hidden Markov model (HMM) trained on bubble trajectories to identify a characteristic U-shaped velocity profile: a transition from high-velocity arteriolar inflow, through a low-velocity capillary transit, to high-velocity venous outflow. Dwell time is defined as the duration the HMM assigns to the low-velocity capillary state, and a trajectory is counted as a complete capillary track when the full arteriole-capillary-venule sequence is recovered.



\section{Results}
\label{sec:results}

\subsection{AER balances high spatiotemporal coherent signals to noise}

Because SVD filtering reconstructs the temporal stack as a linear combination of a subset of eigenspaces, the SVD-filtered result may still retain conventional speckle noise, whereas in the whitened data, where the tissue subspace has been suppressed by rescaling its eigenvalues to the noise floor, the residual noise and tissue signal is white (see Supplementary Video 1). This is also reflected in the covariance domain (Fig.~\ref{fig:qualitative_covariance}, bottom): after SVD filtering, in which the large eigenvalues corresponding to highly coherent tissue signal are removed, the covariance retains a strong diagonal but exhibits residual off-diagonal spatiotemporal correlations, whereas the covariance of the whitened dataset closely approximates the identity, as expected from Equation~\ref{eq:identity}, indicating that each spatiotemporal index is uncorrelated. This decorrelation is directly reflected in the single-frame image quality shown in Fig.~\ref{fig:qualitative_covariance} (top).

To build intuition for how AER reshapes the signal at the pixel level, we plot the in-phase (I) against the quadrature (Q) component over time at three representative points (Fig.~\ref{fig:real_imag}): a stationary, temporally coherent signal, such as skull or tissue, traces an elongated, off-center scatter, since successive I and Q samples are correlated, while a fully decorrelated signal produces a circular cloud centered at zero. At the skull (first row), SVD filtering partially recenters the distribution but leaves a directional, correlated structure, whereas ZCA removes it entirely. The same pattern holds at a large cortical vessel (second row); at a large artery near the base of the brain (third row), SVD and ZCA perform similarly, likely because the signal there already approximates a decorrelated distribution before filtering. This decorrelation is particularly beneficial for slow microbubble flow, which shares clutter's temporal coherence and is therefore easily discarded by rank-based filters like SVD; because AER rescales rather than discards eigenmodes, these weak, slow-moving signals are more likely to be preserved.

\subsection{ZCA recovers flow across the full eigenvalue spectrum}

\begin{figure}[!t]
    \centering
    \includegraphics[width=\linewidth]{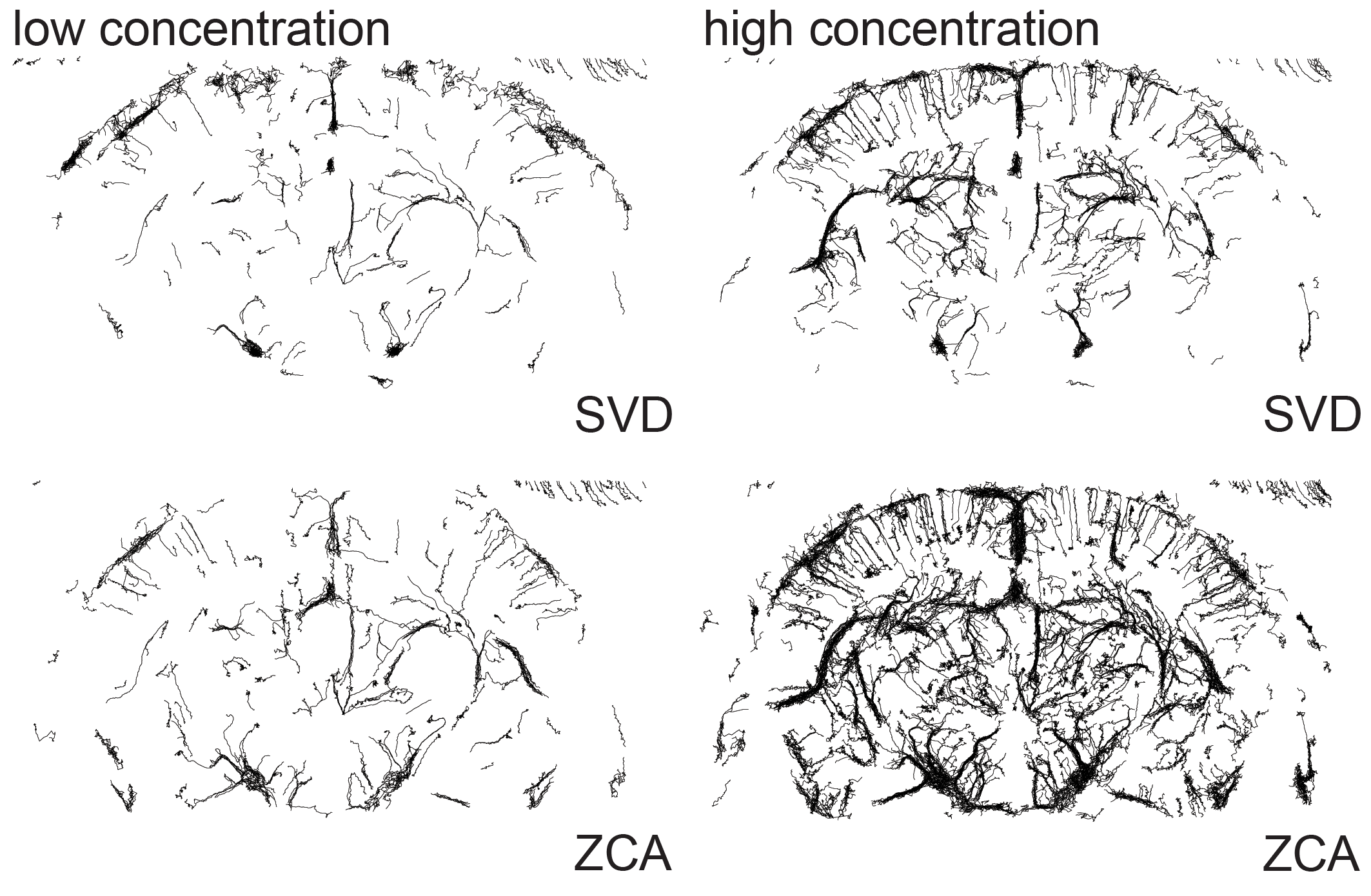}
    \caption{Representative tracking results on 5 seconds of microbubble data at periods of low and high concentration regimes within the same mouse dataset. low concentration (left) and high concentration (right) are compared between SVD (top) and ZCA (bottom). All tracking was performed with the same parameters and radial symmetry localization.}
    \label{fig:tracking}
\end{figure}

\begin{figure}[!t]
    \centering
    \includegraphics[width=1.0\linewidth]{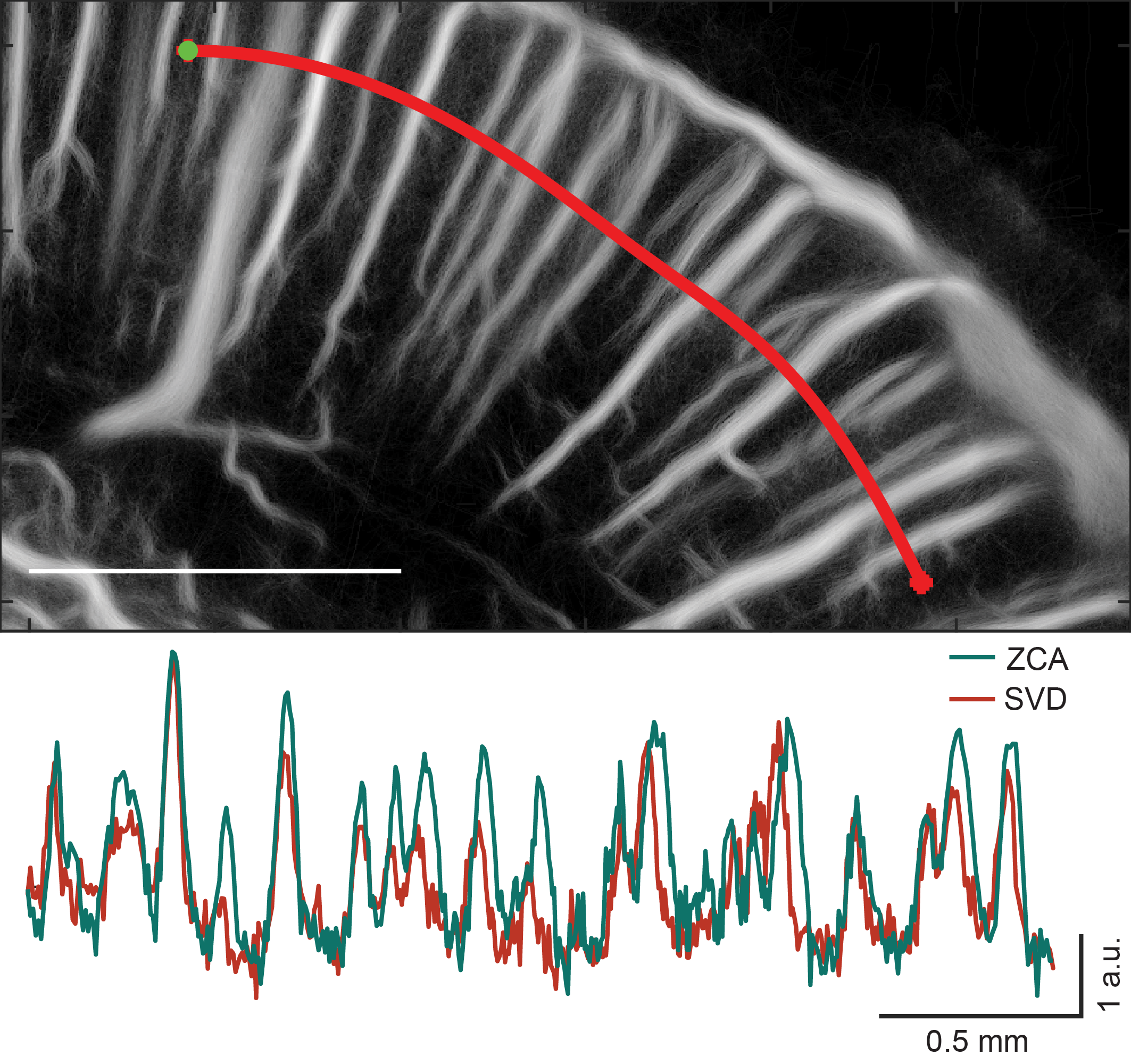}
    \caption{Cortical microvascular reconstruction vessel profile, cropped from the same SVD/ZCA reconstructions shown at full-brain scale in Fig.~\ref{fig:ULM} below and rendered in grayscale, rather than the colormap of Fig.~\ref{fig:ULM}, to improve visibility of the traced profile line. Image density profiles across the red line for SVD and ZCA reconstructions of a 36 minutes scan.}
    \label{fig:vessel_profile}
\end{figure}

\begin{figure*}[!t]
    \centering
    \includegraphics[width=1.0\linewidth]{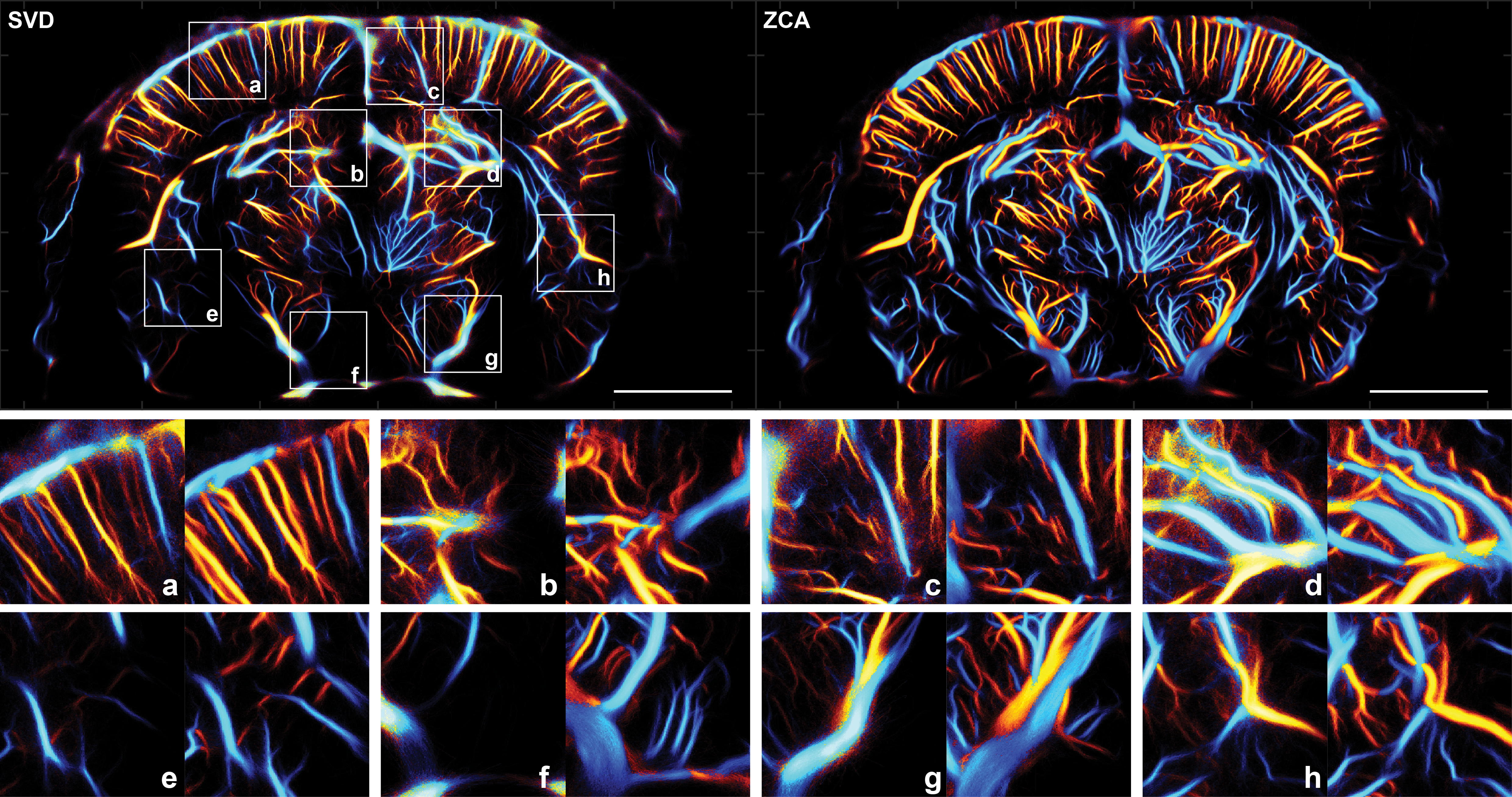}
    \caption{Qualitative ULM results after SVD clutter filtering and ZCA in the temporal domain for a 36 minute continuous infusion acquisition. Sub-figures a-h (middle) are overlaid on the full SVD image (top). (Bottom left) Representative velocity maps for SVD and ZCA. (Bottom right) vessel profiles along the cortex evaluated for both conditions.}
    \label{fig:ULM}
\end{figure*}

\begin{figure}[!t]
    \centering
    \includegraphics[width=1.0\linewidth]{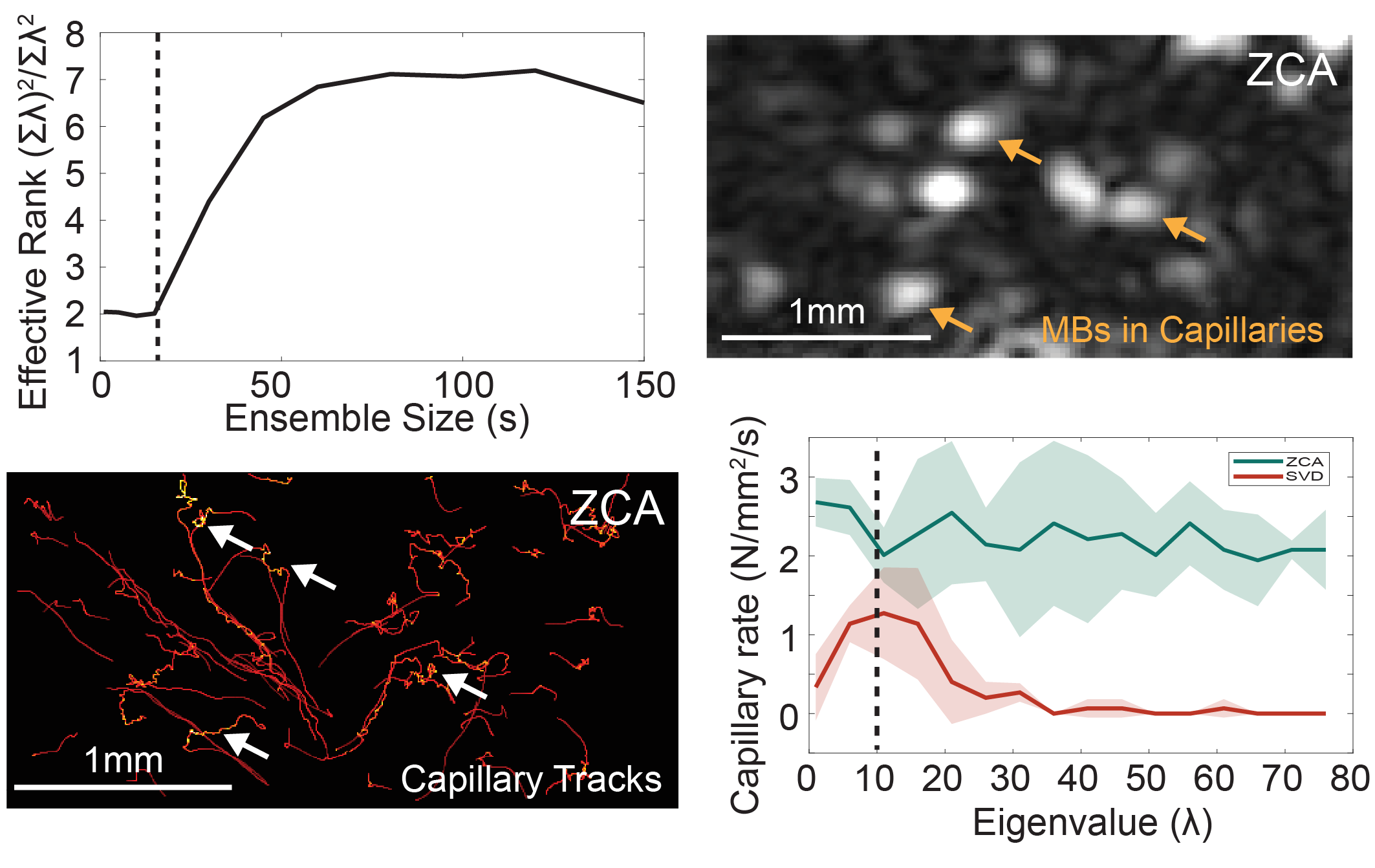}
    \caption{Parameter sweep and capillary trajectory optimization. (top) Effective rank of the SVD decomposition as a function of temporal ensemble size (vertical line at 15 s) and characteristic Bmode frame of stationary microbubbles (yellow arrows). (bottom) Characteristic capillary tracking results of full arteriovenous transits and optimization of capillary trajectory detection as a function of temporal ensemble size (vertical line at 10 $\lambda$).}
    \label{fig:velocity}
\end{figure}

As described in Equation \ref{eq:reg}, the ZCA weighting includes a regularization parameter $\lambda$ to stabilize the matrix inversion, introducing a free parameter that may influence clutter suppression performance: setting $\lambda = \lambda_1$ yields no filtering, whereas setting $\lambda = \lambda_n$ whitens all eigenspaces to the noise floor of the dataset, and varying $\lambda$ modulates the texture of the output from speckle noise toward white noise. We therefore characterized the reconstructed dataset as a function of $\lambda$, sweeping it from 1 to 1000 (corresponding to 1 second of data at 1000 fps) for a single RF buffer and estimating clutter, flow, and noise power as described in Section~\ref{sec:param_opt}; in the SVD case, $\lambda$ plays the equivalent role of the eigenvalue cutoff used for flow reconstruction. Fig.~\ref{fig:ens_compare} (top) shows the resulting normalized power for ZCA (solid lines) and SVD (dotted lines).

According to Fig.~\ref{fig:ens_compare}, the clutter signal drops off immediately in both cases, though more gradually for ZCA, while flow and noise both increase with $\lambda$; noise eventually overtakes flow, but this crossover occurs later for ZCA than for SVD. The largest difference, however, is that flow is recovered across the entire eigenvalue spectrum for ZCA, whereas it is heavily attenuated in SVD outside a narrow range. Fig.~\ref{fig:ens_compare} (bottom) shows representative reconstructions for each Fourier band at $\lambda = 1$, $26$, and $1000$ to visualize the performance of the clutter filtering across all $\lambda$ values.

This qualitative comparison illustrates the same divergence directly. At $\lambda=1$, SVD and ZCA are visually indistinguishable across all three bands, consistent with $\lambda=\lambda_1$ corresponding to no filtering. At $\lambda=26$, near the peak of the flow curves, both methods recover a well-defined vascular pattern in the flow band with comparably suppressed clutter. At $\lambda=1000$, however, the two methods diverge sharply: for SVD, the clutter band has collapsed to a flat, saturated value with no discernible structure and the flow band has been reduced to unstructured noise, while for ZCA a faint clutter pattern remains visible and, more notably, the vascular flow pattern is still clearly resolved despite the added noise texture. This confirms qualitatively that flow information collapses under heavy SVD regularization while it is preserved, at the cost of added noise, under heavy ZCA whitening. In AER terms (Equation~\ref{eq:aer}), increasing $\lambda$ gives every flow mode past the SVD cutoff zero weight, whereas ZCA only lowers the rescaling floor, so all flow modes keep a nonzero weight.

In practice, this indicates that ZCA is more forgiving to the choice of regularizer than SVD: a single $\lambda$ can be reused across acquisitions that differ in microbubble concentration, imaging depth, or skull thickness, without re-tuning for each dataset, and flow reconstruction remains stable even as clutter and noise levels drift over a long acquisition.

\begin{figure*}[!t]
    \centering
    \includegraphics[width=1.0\linewidth]{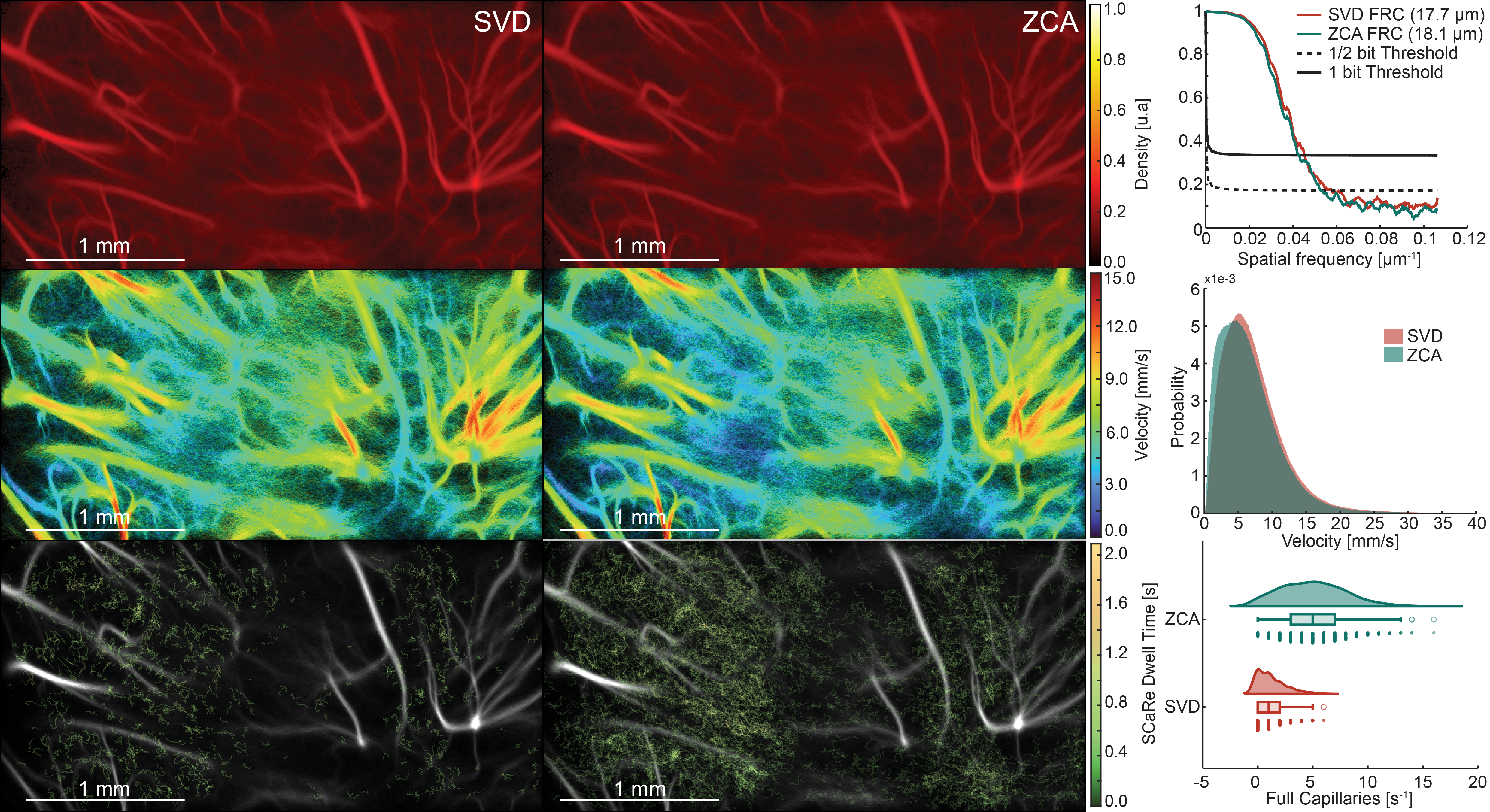}
    \caption{Representative ULM tracking results on subcortical ROI. (top) ULM density maps for SVD and ZCA reconstructions. FRC estimates with 1/2 bit and 1 bit Thresholds and estimated resolutions. (middle) ULM velocity projections for SVD and ZCA with associated histograms on the right. (bottom) SCaRe reconstruction maps for SVD vs ZCA and violin rainfall plot of full capillaries recovered per second.}
    \label{fig:capillaries}
\end{figure*}

\subsection{ZCA increases the number of trajectories}

We can view the effects of ZCA on the temporal covariance in the tracking behavior on the same dataset across low and high concentration regimes (Fig.~\ref{fig:tracking}). In low concentration, ZCA recovers more trajectories across the whole field of view, indicating that attenuation is inherently compensated, and tracking artifacts concentrated at the skull-brain interface under SVD are visibly reduced. The same trend holds, and is further enhanced, at high concentration: for the same amount of data, ZCA recovers more trajectories, suggesting that comparable ULM maps could be constructed with less data.

This difference is most apparent in the central and peripheral vascular arcs of Fig.~\ref{fig:tracking}: under SVD, only fragments of the arborizing network are recovered, with visible gaps between disconnected trajectory segments, whereas under ZCA the same field of view shows a near-continuous, fully arborized network. The improvement is most pronounced in the high-concentration case, where SVD still leaves gaps at the periphery that ZCA fills in almost completely.

This increase in trajectory density is directly visible in the resulting cortical vasculature (Fig.~\ref{fig:vessel_profile}). On large vessels like the vein, both methods yield the same vascular profile, indicating full reconstruction, but the ZCA profile line shows greater definition in smaller penetrating arterioles, along with new vasculature separable from noise and a shift in the estimated vessel center. This is consistent with SVD's vasculature being underpopulated due to its lower trajectory count, though without ground truth we cannot rule out that the discrepancy instead reflects a bias introduced by ZCA's rescaling of coherent eigenspaces. Overall, ZCA populates this cortical region more fully than SVD, with equivalent or better reconstruction in highly dense vascular territories.

The same gain in trajectory density and vascular completeness holds at the scale of the entire brain. Fig.~\ref{fig:ULM} shows the complete reconstruction over 36 minutes of continuous acquisition, including the cortical region highlighted in Fig.~\ref{fig:vessel_profile} above: ZCA compensates for attenuation, more completely reconstructing vasculature at the bottom of the cortex and subcortex, with clearer patterns in high-density zoom regions (a-d) and fuller vasculature elsewhere (e-h). ZCA also suppresses skull clutter, seen as reduced tracking artifacts above the cortical surface, and enhances the large feeding and draining arteries and veins at the bottom of the brain (f-g), which are normally thinned by aliasing in SVD-filtered data.

\subsection{Spatial covariance AER enables longer ensembles to recover slow flow and capillary tracks}

Results up until now have focused on vessel reconstruction quality, but capillary-level flow is not resolved in Fig.~\ref{fig:ULM}: forming the $N_t \times N_t$ temporal covariance matrix over long ensembles is memory-limited at the beamforming density used here (four points per wavelength), which was retained for image quality rather than reduced to permit longer ensembles. Increasing the temporal statistics inside the matrix decomposition resolves these stationary microbubbles, revealing blood transit through the entire arteriovenous tree \cite{lee2026assessing}. Operating on the spatial rather than temporal covariance keeps this tractable over arbitrarily long ensembles without growing the covariance matrix size. Here we explore a $64 \times 192$ subcortical ROI; because this analysis uses a different covariance formulation and a much smaller field of view than Fig.~\ref{fig:ULM}, the following figures use different intensity and velocity colormaps than the whole-brain visualization above.

To estimate the optimal ensemble size, we measured the effective rank of the SVD decomposition as a function of temporal samples (Section~\ref{sec:param_opt}); Fig.~\ref{fig:velocity} (top left) shows it increasing sharply before plateauing at 15~s, the ensemble length used for the remainder of this section. The additional temporal statistics reveal, in both the filtered and tracking data, stationary bubbles in areas of dense microvasculature that remain separable from faster-moving scatterers, along with the arteriole-to-capillary-to-venule trajectories underlying the Single Capillary Reporter (SCaRe) metric (Section~\ref{sec:scare}) \cite{lee2026assessing}.

Using SCaRe, we measured the number of full capillary tracks recovered by ZCA- and SVD-filtered data as a function of $\lambda$, mirroring the stable behavior seen in Fig.~\ref{fig:ens_compare}: a high-pass filter initially increases capillary counts as stationary clutter is removed, then reduces them as spatiotemporal artifacts appear; the count per area per second remains stable across $\lambda$ from 1 to 80, with ZCA and SVD converging to nearly the same value, though ZCA consistently records slightly more trajectories. Comparing density, velocity, and SCaRe-derived distributions for the same subcortical dataset (Fig.~\ref{fig:capillaries}), density reconstructions show comparable resolution (FRC: 17.7~$\mu$m SVD, 18.1~$\mu$m ZCA), while ZCA's velocity distribution extends further into the low-velocity range and yields longer capillary dwell times and a higher rate of fully connected capillary tracks: 1 capillary/s for SVD versus 5/s for ZCA (max 20/s), with the reconstructed capillary population correlating with low-velocity regions on the projected velocity maps.

\section{Discussion}

Here we present adaptive eigenspectrum remapping (AER), instantiated as spatiotemporal ZCA whitening, as an alternative to SVD-based clutter filters for transcranial brain imaging and ULM. By removing correlations between spatiotemporal eigenspaces, AER with ZCA weighting enhances microbubble reconstruction and improves skull clutter suppression at large ensemble sizes. Its most significant advantage is eliminating the manually tunable eigenvalue cutoff, which is still frequently implemented by hand despite proposed automated methods \cite{baranger2018adaptive}, since the SVD is data-dependent and different acquisition configurations or anatomical targets require different cutoffs. Rather than selecting one, AER balances the spatiotemporal eigenspaces, rescaling highly coherent signals such as skull tissue toward the noise floor, generalizing the spectral flattening of the Wiener-Khinchin theorem \cite{papoulis2002probability} to the spatiotemporal domain while retaining the full signal.

Because no data are discarded, reconstruction quality remains consistent across ensemble lengths and ZCA avoids SVD-related artifacts such as aliasing of high-velocity flow; unlike PCA whitening, it also preserves the original data space through back-projection, retaining information about flow directions. AER can be implemented before or after beamforming, on spatial or temporal covariance matrices, whichever is more convenient, and is broadly applicable wherever spatiotemporal data contain distinct flow regimes, particularly as gapless acquisition makes long temporal sequences increasingly practical.

We framed this work under the broader AER term because whitening is only one choice of eigenvalue weights. Complete whitening flattens the spectrum to the noise floor, which our $\lambda$ sweep shows suppresses clutter but also adds noise texture at heavy regularization. Other weightings may be preferable in some settings, for example partial whitening that compresses rather than flattens the clutter eigenvalues, shrinkage-type weights that leave the noise floor unamplified, or weights informed by the temporal frequency content of each eigenvector to emphasize slow or fast flow regimes selectively. Because these all share the same eigendecomposition and back-projection (Equation~\ref{eq:aer}), they can be compared within one framework at no additional computational cost. We leave the systematic comparison of AER weightings to future work.

Several limitations remain: given the diversity of ultrasound applications, it remains to be established whether AER performs reliably under all conditions, and in microvascular regimes with significant tissue motion, such as myocardial imaging, estimating the eigenvector subspaces becomes more challenging, though we expect dominant coherent structures can still be identified with sufficient temporal sampling. Numerical precision is a further consideration, since the ZCA weighting's inverse square root amplifies estimation error and floating-point round-off for small eigenvalues; Tikhonov regularization limits this by replacing $\lambda_k^{-1/2}$ with $(\lambda_k+\lambda)^{-1/2}$, at the cost of a controlled deviation from exact whitening, and the regularization level should be reported since it governs the balance between clutter suppression, flow preservation, and noise amplification.

A separate consideration is the choice of decomposition: preliminary analysis suggests randomized SVD still requires accurate estimation of the full eigenvalue spectrum, making full subspace estimation preferable for AER. Future work should investigate whether a single transformation matrix $W$ can be estimated \textit{a priori} for an entire dataset and applied in real time.

\section{Conclusions}
In this work, we introduced adaptive eigenspectrum remapping (AER) and investigated ZCA matrix whitening as an AER weighting and an alternative to hard-thresholded SVD clutter filtering for transcranial microbubble imaging and ULM in the mouse brain. By rescaling the covariance eigenspectrum rather than discarding selected modes, ZCA reduces the dominance of highly coherent clutter while retaining weak and slow-flow components. Across temporal and spatial covariance formulations, ZCA improved microbubble contrast and vascular image quality, produced higher trajectory density, and revealed more low-velocity flow in ULM reconstructions. With sufficiently long temporal ensembles, spatial AER maintained stable reconstructions, while SCaRe analysis showed more complete capillary networks and connected artery-to-capillary-to-vein trajectories than SVD. These results support AER as a flexible, data-adaptive alternative to manually selected SVD cutoffs, although its performance remains dependent on covariance estimation, numerical precision, and the choice of regularization. More broadly, the results suggest that remapping the covariance eigenspectrum, of which whitening is one instance, can improve microvascular reconstruction while preserving information that threshold-based clutter filtering may remove.

\bibliographystyle{IEEEtran}
\bibliography{bibliography.bib}

\end{document}